Pressure Evolution of Magnetism in URhGa

M.Míšek[1], P. Opletal[2], V. Sechovský[2], J. Kaštil[1], J. Kamarád[1], M. Žáček[1,2] and J.Prokleška[2]

[1]Institute of Physics, Academy of Sciences of Czech Republic, v.v.i, Na Slovance 2, 182 21, Prague 8, Czech Republic
[2]Charles University, Faculty of Mathematics and Physics, Department of Condensed Matter Physics, Ke Karlovu 5, 121 16 Prague 2, Czech Republic

*In this paper, we report the results of an ambient and high pressure study of a 5f-electron ferromagnet URhGa. The work is focused on measurements of magnetic and thermodynamic properties of a single crystal sample and on the construction of the p-T phase diagram. Diamond anvil cells were employed to measure the magnetization and electrical resistivity pressures up to ~ 9 GPa.*
*At ambient pressure, URhGa exhibits collinear ferromagnetic ordering of uranium magnetic moments $\mu_U$ ~ 1.1 $\mu_B$ (at 2 K) aligned along the c-axis of the hexagonal crystal structure below the Curie temperature $T_C$ = 41K. With the application of pressure up to 5GPa the ordering temperature $T_C$ initially increases whereas the saturated moment slightly decreases. The rather unexpected evolution is put in the context of the UTX family of compounds.*

I. INTRODUCTION

URhGa belongs to the group of U*TX* (*T* - transition metal, *X* - p-element) compounds, crystallizing in the hexagonal ZrNiAl - type structure (P-6m2). These uranium intermetallic compounds are formed for the late transition metals (*T*) and the III-V group p-elements (*X*) [1]. Magnetism in these intermetallics is governed by two key mechanisms – overlap of 5f wave functions of the neighboring uranium atoms and hybridization of the uranium 5f states with the valence states of *T* and *X* ligands. The very strong uniaxial magnetic anisotropy locks the U magnetic moments aligned along the c-axis of the hexagonal structure whereas only Pauli paramagnetism is observed in the perpendicular directions.

Several these compounds order ferromagnetically providing a playground for investigation of the critical behavior of itinerant ferromagnets [2-5]. Contrary to the case of antiferromagnetism where novel states are result of the presence of quantum critical point (QCP; see e.g. [6],[7]), the ferromagnet-to-paramagnet quantum phase transitions (QPT) have been found to be the cause of these phenomena opening the whole new chapter of investigation of ferromagnetic materials ranging from the classical and quantum description of low temperature order-disorder transitions to fundamental question of the existence of ferromagnetic QCP [8].

Particularly, the pressure-induced ferromagnet-to-paramagnet transition at low temperatures has been discussed within different scenarios, where quantum phase transitions may take place. In sufficiently clean systems at low temperatures, there are two possible scenarios: Either a first-order ferromagnetic-to-paramagnetic transition, or the appearance of an inhomogeneous magnetic phase between the ferromagnetic and paramagnetic state [8]. There are only few known U-based ferromagnets displaying the change of the order of transition, demonstrating the appearance of the first order transition by driving the system close to the loss of the long range magnetic order [8].

Up to now the URhGa compound has been characterized only at ambient pressure and in polycrystalline/powder form. In this paper we report the preparation of a single crystal, compare its properties to previously published data and study the evolution of magnetism under pressure, comparing it to the neighboring UCoGa and discuss the peculiar behavior in the context of the entire U*TX* series.

II. EXPERIMENTAL

A precursor for single crystal growth was prepared by arc melting using high-quality elements U (SSE purified), Rh(99.95%) and Ga (99.9999%). The single crystals has been grown using Czochralski method in a triarc furnace with pulling speed 12 mm/hr. The product was wrapped in a Ta foil and sealed in a quartz tube evacuated to $10^{-7}$ mbar. The annealing was done at 900°C for 3 weeks. The structure and composition of the single crystal was checked by X-ray Laue method (Laue diffractometer made by Photonic Science), X-ray powder diffraction (XRPD, Bruker AXS D8 Advance X-ray difractometer with Cu X-ray tube) and energy dispersive X-ray spectroscopy (EDX, scanning electron microscope Tescan Mira I LMH). Samples for specific experiments were cut by a wire saw from the single crystal oriented with a Laue diffractometer. Magnetic measurements were performed using a MPMS 7 XL SQUID based magnetometer (Quantum Design) with the magnetic field applied along the c-axis. For the application of pressure, the "turnbuckle"- type diamond anvil cell [9] was used. Specific heat was measured using a PPMS apparatus (Quantum Design).

III. RESULTS

The measurement of heat capacity show a clear transition at $T_C = 41$ K being quickly suppressed by applied magnetic field along the c-axis. The magnetization study at low temperatures revealed a ferromagnetic ground state (see Fig 2 and Fig 1 inset) with ordered moment of 1.1 $u_B$/f.u. for field applied along the c-axis. The magnetic response for field applied within the basal plane remains paramagnetic, similar to other members of the family. Above the ordering temperature the evolution of magnetic susceptibility is field independent and follows Curie-Weiss law with effective moment 2.45 $\mu_B$/f.u. and paramagnetic Curie temperature 45 K (see the inset of Fig 2).

The above mentioned parameters qualify the compound as a suitable candidate for the possible search of scaling laws close the ferromagnetic quantum criticality with UCoGa compound being the benchmark partner (same structure and magnetic order, similar ordering temperatures, nevertheless half ordered moment) [3, 8]. In order to test this scenario a pressure study of magnetic properties has been made. Temperature dependencies of magnetization at small fields (0.01 T) measured at selected pressures are depicted in Fig 3. With increasing pressure, the transition remains visible, however we clearly see that contrary to expectation the Curie temperature increases with increasing pressure and reaches the plateau in the range of 5-8 GPa. As the temperature dependencies indicate and field scans at low temperatures confirm (Fig. 4), the ground state becomes more robust (in the sense of increased coercive field), see Fig 4. Contrary, the saturated moment is partly reduced (-0.05 $u_B$/f.u./GPa).

## IV. DISCUSSION AND CONCLUSIONS

The obtained results can be summarized in the proposed magnetic phase diagram as shown in Fig. 5. The magnetic measurements reveal continuous increase of transition temperature up to 48 K in 6 GPa, followed by a plateau up to the highest pressure used in this study (~8.5 GPa). This behavior is rather rare in the studied isostructural U$TX$ family of compounds as the majority of published pressure studies report monotonous suppression of the ordering temperatures down to the presumable quantum phase transition. The sole exception is UPtAl where similar increase Curie temperature has been observed in low pressures [10].

The reasoning for such behavior is unclear. Within one ($T_C$ increasing) or another ($T_C$ decreasing) scenario the situation is rather understandable and possible to explain on the basis of various hybridization effects with d- and p- electrons (see detailed discussion in [1]). However, the key question remains – what is the mechanism behind/conditions for the change of initial polarity of pressure influence on magnetism. Our study shows, that UPtAl is not an isolated anomaly and the border lie on the UCoGa/URhGa and URhAl/UPtAl line. It is highly likely that the application of hydrostatic pressure may have a multiplicative effect - as discussed in Introduction, U$TX$ have layered ZrNiAl-type structure, being strongly anisotropic, both in magnetic and mechanical properties. The directions within basal plane sheets are soft (in terms of compressibility) and shortest U-U distance is within the sheets as well leading to the synergy effect under applied hydrostatic pressure – the closer the uranium atoms are the larger is effect of the pressure (for details see e.g. [1] and references therein). This is usually prevailing effect of hydrostatic pressure and for moderate U-U distances it does not change within the used range of pressures, consequently allowing to use the proportionality between the pressure, tuning of electronic structure and control parameter. However, our study, and the case of UPtAl, shows that this may not be always the case and the origin of this phenomena deserves more detailed studies, both on experimental and theoretical level.

## ACKNOWLEDGMENTS

This work is part of the research program GACR 16-06422S which is financed by the Czech Science Foundation. Experiments were performed in the Materials Growth and Measurement Laboratory MGML (see: http://mgml.eu).

[5] N. Kimura, N. Kabeya, H. Aoki, K. Ohyama, M. Maeda, H. Fujii, M. Kogure, T. Asai, T. Komatsubara, T. Yamamura, I. Satoh, Quantum critical point and unusual phase diagram in the itinerant-electron metamagnet UCoAl, Physical Review B, 92 (2015) 035106.
[6] P. Coleman, A.J. Schofield, Quantum criticality, Nature, 433 (2005) 226-229.
[7] A. Schröder, G. Aeppli, R. Coldea, M. Adams, O. Stockert, H.v. Löhneysen, E. Bucher, R. Ramazashvili, P. Coleman, Onset of antiferromagnetism in heavy-fermion metals, Nature, 407 (2000) 351.
[8] M. Brando, D. Belitz, F.M. Grosche, T.R. Kirkpatrick, Metallic quantum ferromagnets, Reviews of Modern Physics, 88 (2016) 025006.
[9] G. Giriat, W. Wang, J.P. Attfield, A.D. Huxley, K.V. Kamenev, Turnbuckle diamond anvil cell for high-pressure measurements in a superconducting quantum interference device magnetometer, Review of Scientific Instruments, 81 (2010) 073905.
[10] F. Honda, T. Eto, G. Oomi, A.V. Andreev, V. Sechovský, N. Takeshita, N. Môri, Collapse of 5 f -Electron Ferromagnetism in UPtAl Under High Pressures, High Pressure Research, 22 (2002) 159-162.

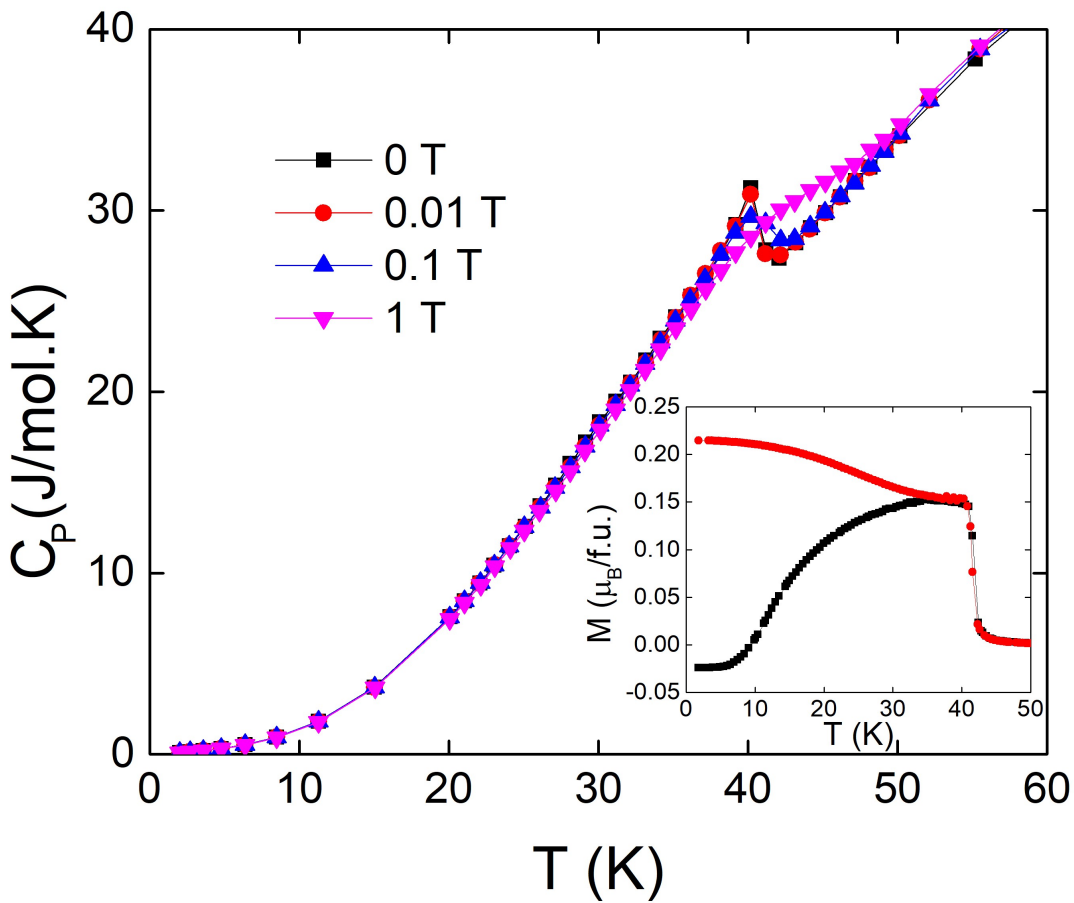

Fig 1 Temperature dependence of specific heat of URhGe measured in various magnetic fields. Inset: Temperature dependence of magnetization of URhGe measured in a field of 0.01 T. ZFC (FC) data represented by black (red) points. Magnetic field applied along the c-axis.

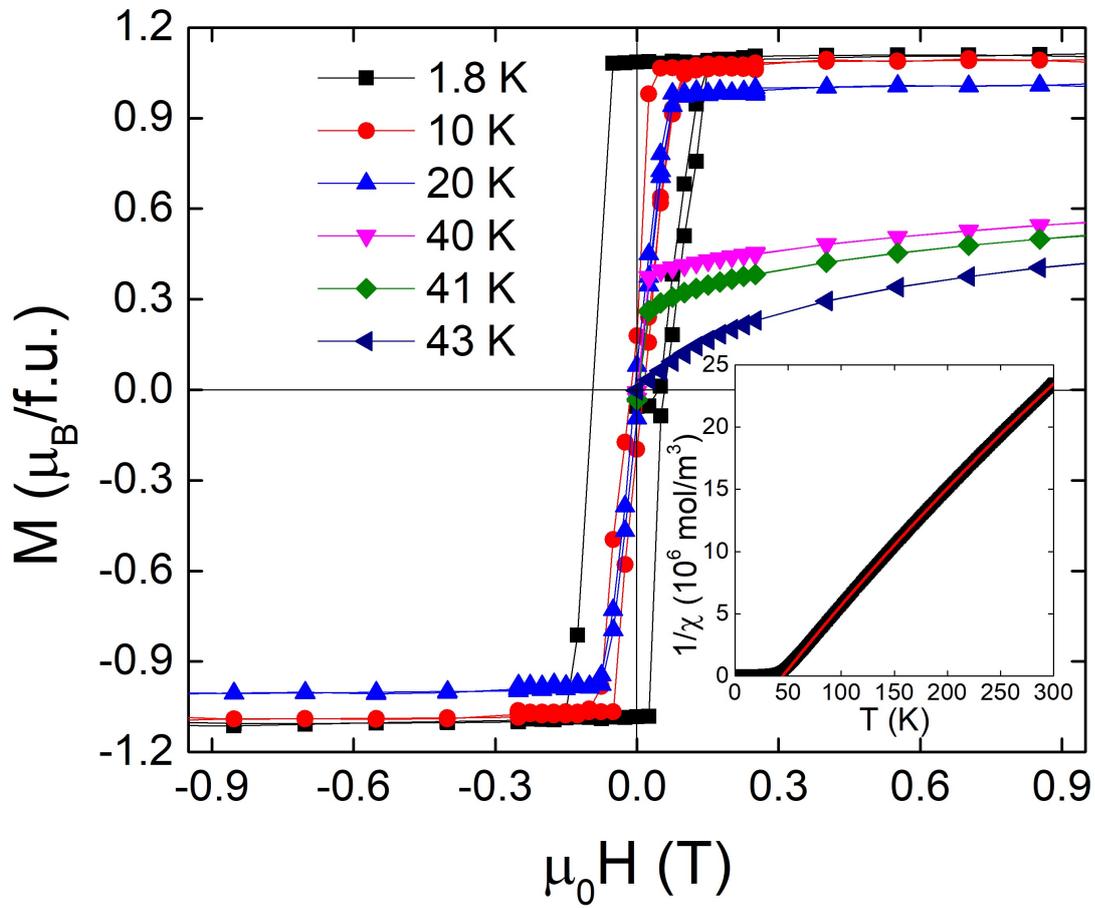

Fig 2 Hysteresis loops of URhGe measured at different temperatures. Inset: Temperature dependence of inverse magnetic susceptibility and a fit (80-300K) to Curie-Weiss law (red line). Magnetic field applied along the c-axis.

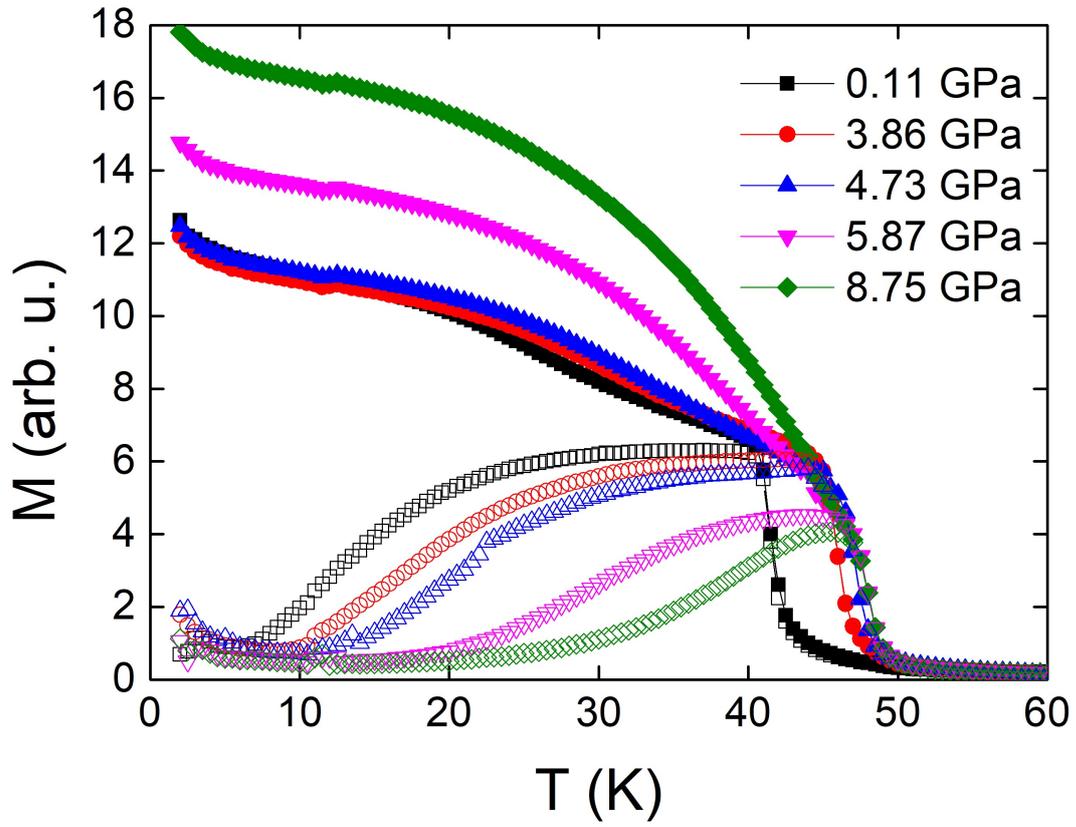

Fig 3. Temperature dependence of magnetization in a field of 0.01 T applied along the c-axis measured at selected pressures.

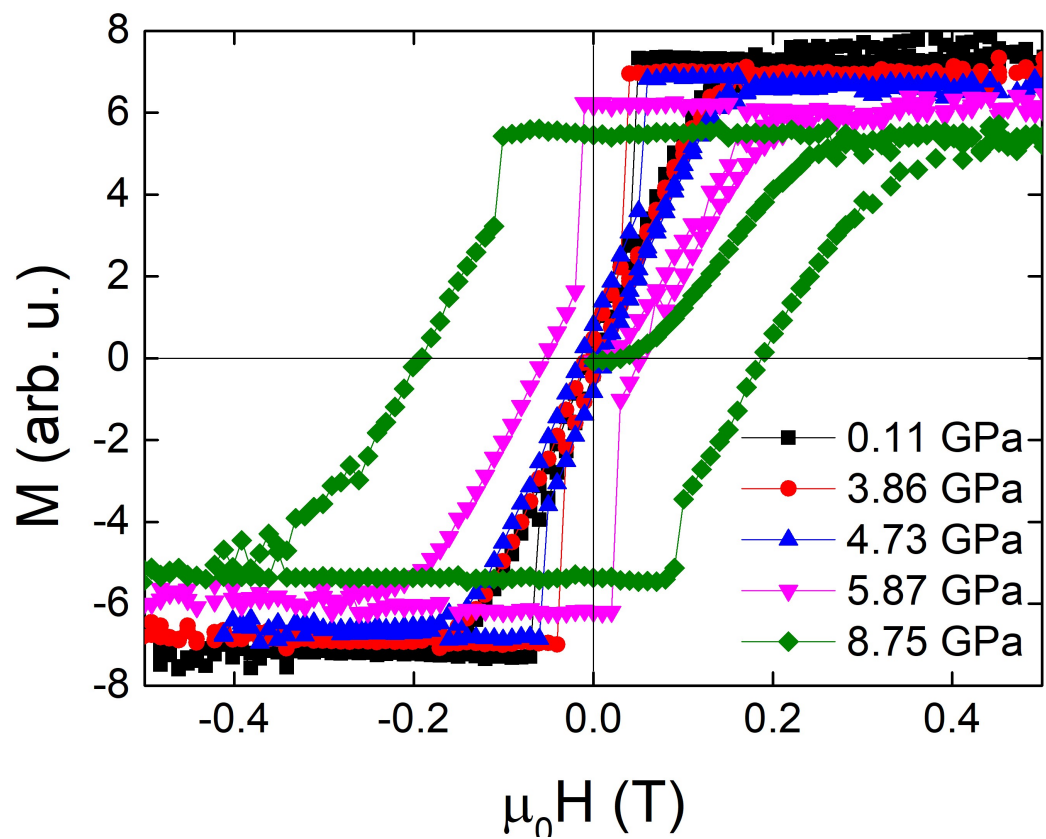

Fig 4. Hysteresis loops of URhGe at 20K measure in selected pressures.

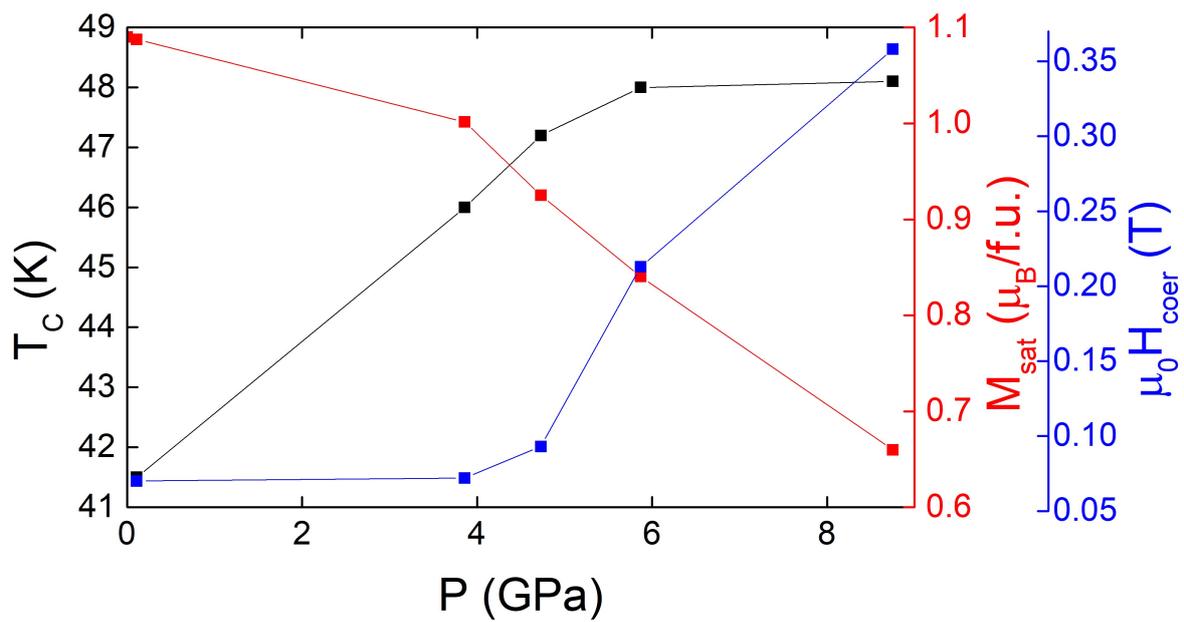

Fig 5. Pressure dependence of ordering temperature, coercive field and saturated magnetic moment (the latter two at $T = 2$ K).